\documentclass[final,3p,times,twocolumn]{elsarticle}

\usepackage{amssymb}
\usepackage{amsmath}
\usepackage{babel}
\usepackage{xcolor}
\usepackage[numbers]{natbib}

\journal{Physica D}

\begin{document}

\begin{frontmatter}



\title{
Universality Classes with Strong Coupling in Conserved Surface Roughening: Explicit vs Emergent Symmetries
}


\author[uc3m,upm]{Pedro Gat{\'o}n-P{\'e}rez} \ead{pedro.gaton.perez@upm.es}
\author[unican,uc3m]{Enrique Rodr\'{\i}guez-Fern\'andez}
\ead{enrique.rodriguez@unican.es}
\author[uc3m]{Rodolfo Cuerno}
\ead{cuerno@math.uc3m.es}

\affiliation[uc3m]{organization={Universidad Carlos III de Madrid, Departamento de Matemáticas and Grupo Interdisciplinar de Sistemas Complejos (GISC)},              addressline={Avenida de la Universidad 30},       city={Leganes},
postcode={28911},
state={Madrid},
country={Spain}}

\affiliation[upm]{organization={ETSIAE Universidad Politécnica de Madrid},
addressline={Plaza Cardenal Cisneros 3},
city={Madrid},
postcode={28040},
state={Madrid},
country={Spain}}

\affiliation[unican]{organization={Departamento de Matemática Aplicada y Ciencias de la Computación, Universidad de Cantabria},
addressline={Bulevard Ronda Rufino Peon 254}, city={Torrelavega},postcode={39316}, state={Cantabria},
country={Spain}}







\begin{abstract}
The occurrence of strong coupling or nonlinear scaling behavior for kinetically rough interfaces whose dynamics are conserved, but not necessarily variational, remains to be fully understood. Here we formulate and study a family of conserved stochastic evolution equations for one-dimensional interfaces, whose nonlinearity depends on a parameter $n$, thus generalizing that of the stochastic Burgers equation, whose behavior is retrieved for $n=0$. This family of equations includes as particular instances a stochastic porous medium equation and other continuum models relevant to various hard and soft condensed matter systems. We perform a one-loop dynamical renormalization group analysis of the equations, which contemplates strong coupling scaling exponents that depend on the value of $n$ and may or may not imply vertex renormalization. 
These analytical expectations are contrasted with explicit numerical simulations of the equations with $n=1,2,$ and 3. For odd $n$, numerical stability issues have required us to generalize the scheme originally proposed for $n=0$ by T.\ Sasamoto and H.\ Spohn [J.\ Stat.\ Phys.\ {\bf 137}, 917 (2009)]. Precisely for $n=1$ and 3, and at variance with the $n=0$ and 2 cases (whose numerical exponents are consistent with non-renormalization of the vertex), numerical strong coupling exponent values are obtained which suggest vertex renormalization, akin to that reported for the celebrated conserved Kardar-Parisi-Zhang (cKPZ) equation. We also study numerically the statistics of height fluctuations, whose probability distribution function turns out (at variance with cKPZ) to have zero skewness for long times and at saturation, irrespective of the value of $n$. However, the kurtosis is non-Gaussian, further supporting the conclusion on strong coupling asymptotic behavior. The zero skewness seems related with space symmetries of the $n=0$ and 2 equations, and with an emergent symmetry at the strong coupling fixed point for odd values of $n$.
\end{abstract}




\end{frontmatter}



\section{Introduction}\label{sec:intro}

Surface kinetic roughening \cite{Barabasi95,Krug1997} is a well-known instance of critical dynamics far from equilibrium \cite{Tauber2014} that remains of a high current interest. Indeed, beyond its original domain of applicability to the dynamics of fronts or interfaces of e.g.\ thin films, epitaxial systems, or bacterial colonies, kinetic roughening is recently proving of relevance to novel contexts, such as e.g.\ active \cite{Ramaswamy2018,Caballero2018b,Caballero2020} and quantum \cite{Fontaine2022,Wei2022} matter, and even to non-interfacial systems, such as e.g.\ the kinetics of chemical reactions \cite{Mondal2022} or the synchronization of phase or limit cycle \cite{,Gutierrez2023b,Gutierrez2025} oscillators. 

A standard classification of kinetic roughening systems \cite{Barabasi95,Krug1997} proceeds by analogy with that made of equilibrium critical dynamics with respect to the conservation of e.g.\ the total magnetization, in terms of which the well-known models A and B (corresponding to non-conserved and conserved dynamics, respectively) were established \cite{Hohenberg1977,Tauber2014}. Thus, beyond Gaussian (linear) models, for non-conserved kinetic roughening the paramount nonlinear universality class corresponds to the strong coupling behavior of the Kardar-Parisi-Zhang (KPZ) equation \cite{Kardar1986},
\begin{align}
    \partial_t h = \nu_2 \partial^2_x &h + \frac{\lambda_2}{2} (\partial_x h)^2 + \eta(x,t) , \label{eq:kpz} \\
    \langle \eta(x,t) \eta(x',t') \rangle &= D \, \delta(x-x') \delta(t-t') ,
    \nonumber
\end{align}
where $h(x,t)$ describes the height of an interface above position $x\in \mathbb{R}$ along a one-dimensional (1D) substrate at time $t$, $\nu_2, D>0$ and $\lambda_2$ are parameters, and $\eta(x,t)$ is zero-average, Gaussian white noise. The rich strong-coupling (nonlinear) scaling behavior of Eq.\ \eqref{eq:kpz} is being found recently for many different systems (see e.g.\ Ref.\ \cite{Rodriguez-Fernandez2020} and other therein), and it includes not only characteristic values for scaling exponents, but also fluctuation statistics described by the Tracy-Widom (TW) family of probability distribution functions (PDF) \cite{Takeuchi2018}. This family of PDFs, which notably features non-Gaussian skewness and kurtosis values, is providing some sort of generalization of the Gaussian distribution for correlated random variables \cite{Fortin2015}, increasingly applying to a remarkable range of systems and typical scales \cite{makey20}.

For conserved interface dynamics, as relevant e.g.\ under standard Molecular Beam Epitaxy (MBE) growth conditions \cite{Barabasi95,Krug1997}, an analogous role to the KPZ equation is played by the so-called conserved KPZ (cKPZ), nonlinear MBE, or Lai-Das Sarma-Villain equation \cite{Lai1991,Villain1991},
\begin{align}
    \partial_t h = -\nu_4 \partial^4_x &h + \lambda_4 \partial_x^2(\partial_x h)^2 + \eta(x,t) , \label{eq:ckpz}
\end{align}
where $\nu_4>0$ and $\lambda_4$ are parameters, and $\eta$ is henceforth as in Eq.\ \eqref{eq:kpz}. Although analytical solutions are not known for this continuum model in contrast with the 1D KPZ equation \cite{Takeuchi2018}, it is likewise known to feature nonlinear scaling exponent values \cite{Barabasi95,Krug1997,Janssen1981} and non-Gaussian (and non-TW) fluctuation statistics with non-zero skewness and small kurtosis values \cite{Carrasco2016}. 

Very recently, a generalization of the cKPZ equation has been proposed and studied \cite{Caballero2018}, in which an additional term allowed by the symmetry considerations employed in the original derivations of cKPZ \cite{Lai1991,Villain1991} has been added to those already present in Eq.\ \eqref{eq:ckpz}. Although it is not clear to which extent does the ensuing long-time dynamics remain free of instabilities, unambiguous indications of nonlinear behavior are reported \cite{Caballero2018}. In this sense, the study in Ref.\ \cite{Caballero2018} raises an interesting question on the exploration of strong-coupling behavior in conserved surface kinetic roughening which differs from that of the cKPZ equation \cite{Ramaswamy2018}.

In this paper we address this issue from the point of view of still another well-known instance of conserved interface dynamics with non-conserved noise, namely, the stochastic Burgers equation for a scalar field $\phi(x,t)$ (see Ref.\ \cite{Rodriguez-Fernandez2019} and other therein),
\begin{align}
    \partial_t \phi = \nu \partial^2_x & \phi + \lambda \phi \partial_x \phi + \eta(x,t) , \label{eq:burgers}
\end{align}
where $\nu>0$ and $\lambda$ are parameters. Indeed, the deterministic terms on the right hand side of Eq.\ \eqref{eq:burgers} add up to a total space derivative, see below. Although strongly related with the KPZ equation, Eq.\ \eqref{eq:burgers} also features nonlinear scaling exponent values which differ from the KPZ ones \cite{Rodriguez-Fernandez2019}. In turn, if the noise in Eq.\ \eqref{eq:burgers} is conserved, then the exponents do derive from those in the KPZ equation ---as in that case Eq.\ \eqref{eq:burgers} becomes the equation for the slope $\phi=\partial_x h$ implied by a KPZ equation for $h$---  \cite{Forster1977}, but not the height PDF \cite{Rodriguez-Fernandez2020}. Remarkably, the field PDF is Gaussian (and in particular, symmetric) for the nonlinear Eq.\ \eqref{eq:burgers} \cite{Rodriguez22}, in spite of the model not being up-down ($\phi\leftrightarrow -\phi$) symmetric. This property can be traced back to the equation remaining invariant, rather, under a combined inversion-reflection operation $(x,\phi)\leftrightarrow (-x,-\phi)$ \cite{Rodriguez22}.

Our goal is to investigate strong coupling behavior in conserved surface kinetic roughening by addressing the following family of evolution equations, that can be considered as a generalization of Eq.\ \eqref{eq:burgers},
\begin{equation}\label{eq:n}
   \partial_t \phi=- B_n \partial_x^4 \phi + \lambda_n \partial_x (\phi \partial_x^n\phi) + \eta(x,t) ,
\end{equation}
where $B_n>0$ and $\lambda_n$ are parameters, and $n=0,1,2,3,\ldots$ Note that, for any value of $n$, Eq.\ \eqref{eq:n} has the form of a conservation law except for the noise term, namely,
\begin{align}
\partial_t \phi = - &\partial_x J_n(x,t) + \eta(x,t), \; \mbox{where} \label{eq:cons_law} \\
J_n(x,&t) = B_n \partial_x^3 \phi - \lambda_n \phi \partial_x^n \phi . \nonumber
\end{align}
For $n=0$ the nonlinear term in Eq.\ \eqref{eq:n} coincides with that of Eq.\ \eqref{eq:burgers} if $2\lambda_0 = \lambda$. In contrast, note that the linear derivative term in Eq.\ \eqref{eq:burgers} is second order (physically representing smoothing by e.g.\ surface tension \cite{Barabasi95,Krug1997}), while that in Eq.\ \eqref{eq:n} is fourth order (physically representing smoothing by e.g.\ surface diffusion \cite{Barabasi95,Krug1997}). However, as seen in Ref.\ \cite{Rodriguez-Fernandez2019} and below, for $n=0$ renormalization group analysis and numerical simulations yield the same strong coupling exponents for both equations, which hence coincide effectively in their asymptotic behavior.

By formulating different equations as a one-parameter family, we can systematically explore generic large-scale behavior within an unified framework. We will pay particular attention to the three first members (aside from the $n=0$ Burgers equation) of the family with $n=1,2,$ and $3$. As it turns out, parameter conditions can be found for each of these cases in which the long-time behavior is free from morphological instabilities, hence significant conclusions can be reached on the occurrence of strong coupling behavior. Moreover, the nonlinearities that appear in these equations bear physical importance by themselves in classic interfacial contexts like the dynamics of thin fluid films or amplitude equations for pattern forming systems, or in more novel contexts like active matter. 

Indeed, the $n=0$ and the $n=1$ nonlinearities occur in the so-called dissipation modified Korteweg-de Vries equation, an universal amplitude equation for Type II pattern formation in systems constrained by a conservation law \cite{Shklyaev2017}. For fluid systems, the $n=1$ nonlinearity in Eq.\ \eqref{eq:n} is a particular case of the so-called porous medium equation \cite{Ockendon1995} that generalizes the diffusion equation to a system in which the diffusion coefficient depends (linearly in this case) on the diffusing field itself. Both this $n=1$ and the $n=3$ nonlinearities occur very frequently in the weakly nonlinear description of convective transport and capillarity, respectively, in the relaxation of thin (and ultrathin, as in surface nanostructuring by ion-beam irradiation \cite{Munoz-Garcia2019}) viscous fluid films \cite{Oron1997}. For example, they both occur together in a description of density-stratified Hele-Shaw flows \cite{Goldstein1993}. In all these contexts, by far the most frequently studied form of the $n=1$ and the $n=3$ nonlinearities is for 1D systems, e.g.\ 1D interfaces.
More recently, the $n=1$ term, together with the KPZ nonlinearity, competes with the standard terms of the time-dependent Ginzburg-Landau (TDGL) equation in the so-called active model A of active matter systems \cite{Caballero2020}. Taking in turn the space Laplacian, the $n=3$ term appears (competing with the cKPZ nonlinearity and with the terms of the conserved TDGL equation) in the so-called active model B+, also relevant to active matter 
\cite{Tjhung2018,Caballero2018b}. 

While the $n=0$ (Burgers) nonlinearity is a paradigmatic description of e.g.\ fluid transport \cite{Bec2007}, the $n=2$ one occurs crucially in the Sawada-Kotera or Caudrey-Dodd-Gibbon \cite{Sawada1974,Caudrey1976}, soliton bearing equation of Mathematical Physics \cite{Zwillinger2021}, and the nonlinear contribution in the conserved current $J_2$ for $n=2$ in Eq.\ \eqref{eq:cons_law} has also has been identified only very recently in particle diffusion that conserves the system center of mass (inducing a Laplacian in front of $J_2$) \cite{Han2024}. Generally, our even-$n$ equations share the same symmetries with respect to $x$ and $\phi$ of the Burgers ($n=0$) case. Within the model family, Eq.\ \eqref{eq:n}, we take $n=2$ to represent the type of behavior obtained for even values of $n$. We will see that it also induces well-defined strong coupling behavior. 

This paper is organized as follows. After recalling in Sec.\ \ref{sec:scaling} the quantities and properties that will be later employed in the study of Eq.\ \eqref{eq:n}, we will first extract predictions on scaling exponent values through a Dynamic Renormalization Group (DRG) analysis in Sec.\ \ref{sec:analytic}. These predictions will be then compared in Sec.\ \ref{sec:numeric} with results from numerical simulations of Eq.\ \eqref{eq:n} for $n=1,2$, and 3. While the $n=1$ and 2 cases can be dealt with through relatively standard numerical schemes, numerical stability issues have required us to generalize to the $n=3$ nonlinearity the scheme proposed in Ref.\ \cite{Sasamoto09} for the $n=0$ Burgers case. A discussion of our results is presented in Sec.\ \ref{sec:discussion}, which is followed by our conclusions and an outlook in Sec.\ \ref{sec:conclusions}. The paper ends with several appendices covering additional DRG details, the deterministic limit of Eq.\ \eqref{eq:n}, and the numerical schemes employed.

\section{Observables for critical dynamics}\label{sec:scaling}

As already noted, the stochastic Eq.\ \eqref{eq:n} for the scalar ``order parameter'' $\phi(x,t)$ has the form (except for the noise term) of a conservation law for any value of $n$, recall Eq.\ \eqref{eq:cons_law}. Hence, within the context of critical dynamics \cite{Tauber2014}, Eq.\ \eqref{eq:n} prescribes conserved dynamics with non-conserved noise. Under these conditions and, as argued by Grinstein \cite{Grinstein1991,Grinstein1995}, one expects scale-invariant behavior for generic parameter values and all $n$. Hence, the standard observables considered in the study of surface kinetic roughening are expected to be relevant to the study of Eq.\ \eqref{eq:n}.

To characterize fluctuations in a 1D system of size $L$ we use the roughness, $W(t)$, which is the root-mean square deviation,
\begin{equation}\label{eq:rough}
     W(t)=\sqrt{ \Big\langle \frac{1}{L} \int_0^L [\phi(x,t)-\bar{\phi}(t)]^2 dx} \Big\rangle,
\end{equation}
where $\langle\cdot\rangle$ denotes an average over noise realizations and an overbar denotes space average. Interpreting the field $\phi(x,t)$ as describing e.g.\ a moving (1D) surface or interface that starts out initially flat, under kinetic roughening conditions the height values become correlated along the $x$ coordinate so that a correlation length can be defined which increases with time as a power law, $\xi(t) \sim t^{1/z}$, where $z$ is the so-called dynamic exponent \cite{Barabasi95,Krug1997}. In this regime, the roughness also increases with time, as $W(t)\sim t^{\beta}$, where $\beta$ is the growth exponent. Eventually, the correlation length equals the system size $L$, a steady state being reached in which the roughness saturates to a time-independent value $W = W_{\rm sat}\sim L^{\alpha}$, where $\alpha$ is the roughness exponent, related with the fractal dimension of the interface field $\phi(x,t)$ \cite{Barabasi95,Mozo22}. The time required to reach the steady state scales as $t_{\rm sat}\sim L^z$. Consistency of these power laws at $t=t_{\rm sat}$ implies $\beta=\alpha/z$, so that there are only two independent exponents. All this behavior can be recast into the so-called Family-Vicsek (FA) ansatz,
\begin{equation}
    W(L,t) = t^{\beta} f(t/L^z) , \label{eq:wfv}
\end{equation}
where the scaling function $f(u) \sim {\rm const.}$ for $u \ll 1$ and $f(u) \sim u^{-\beta}$ for $u\gg 1$ \cite{Barabasi95,Krug1997}.

A more detailed description of the system is provided by e.g.\ two-point correlation functions. Here, we will address them in Fourier space through the Power Spectral Density (PSD) or structure factor, $S(k,t)$, defined as
\begin{equation}\label{eq:psd}
     S(k,t)= \langle\hat{\phi}(k,t)\hat{\phi}(-k,t)\rangle=\langle{|\hat{\phi}(k,t)|}^2\rangle ,
\end{equation}
where $\hat{\phi}(k,t)$ is the space Fourier transform of $\phi(x,t)-\bar{\phi}(t)$ and $k$ is 1D wave vector. Thus, the magnitude of $S(k,t)$ identifies the relevance of each spatial scale in the surface morphology $\phi(x,t)$ trough its Fourier decomposition. The FV ansatz for the structure factor becomes \cite{Barabasi95,Krug1989}
\begin{equation}
    S(k,t) = k^{-(2\alpha+d)} s(kt^{1/z}), \label{eq:psd_fv}
\end{equation}
where $s(u) \sim u^{2\alpha+1}$ for $u \ll 1$ and $s(u) \sim {\rm const.}$ for $u \gg 1$, and $d=1$ for 1D interfaces. Thus, for distances smaller than the correlation length, $k \gg 1/t^{1/z}$, the structure factor displays power-law correlations as $S(k) \sim 1/k^{2\alpha+1}$, while it is uncorrelated ($S(k) \sim {\rm const.}$) at larger distances where $k \ll 1/t^{1/z}$.

\subsection{Fluctuation statistics}

Beyond scaling exponent values, and originally driven by exact results for the KPZ equation \cite{Takeuchi2018}, recent developments in surface kinetic roughening highlight the fluctuation statistics of the field as another important trait of the universality class, see e.g.\ Refs.\ \cite{Takeuchi2018,Carrasco2016,Rodriguez-Fernandez2019,Rodriguez-Fernandez2020,Rodriguez-Fernandez2021,Rodriguez22} and other therein. Thus, remarkably, the rescaled fluctuation variable
\begin{equation}
\chi = \frac{\phi(x,t)-\bar{\phi}(t)}{W(t)} \label{eq:chi} ,   
\end{equation}
features a stationary PDF which largely characterizes the universality class. While linear models (and some nonlinear models too, notably Eq.\ \eqref{eq:burgers} \cite{Rodriguez-Fernandez2019}) feature Gaussian statistics, deviations from Gaussianity in the field fluctuations become an indication of nonlinear or strong coupling behavior. On the other hand, the scaling exponents may not suffice to completely identify the kinetic roughening universality class. Thus, examples are known of systems that share the same set of exponent values while featuring different PDF or, conversely, systems with different exponents may show the same PDF; see e.g.\ Refs.\ \cite{Barreales2012,Rodriguez-Fernandez2021,Rodriguez22,Marcos2022}.
All of this makes it interesting to assess the fluctuation PDF in a given model.

The fact that the PDF of the rescaled fluctuations, Eq.\ \eqref{eq:chi}, is time-independent, implies in particular that certain cumulant ratios take universal values. Hence, beyond the second cumulant (roughness), it is convenient to also measure the normalized third (skewness) and fourth (kurtosis) cumulants, defined, respectively, as \cite{Takeuchi2018}
\begin{equation}\label{eq:skew}
    {\cal S}(t)=\frac{1}{W^3(t)} \Big\langle \frac{1}{L} \int_0^L [\phi(x,t)-\bar{\phi}(t)]^3 dx \Big\rangle,
\end{equation}
and
\begin{equation}\label{eq:kurt}
    {\cal K}(t)=\frac{ 1}{W^4(t)} \Big\langle \frac{1}{L} \int_0^L [\phi(x,t)-\bar{\phi}(t)]^4 dx \Big\rangle.
\end{equation}
These quantities equal ${\cal S}_{\rm G}=0$ and ${\cal K}_{\rm G}=3$ for a Gaussian fluctuation PDF. In the most general case, they may show a non-trivial time dependence reflecting crossover behavior between the various universality classes influencing the system. E.g.\ for the KPZ equation a non-trivial time evolution occurs from a Gaussian PDF at short times, to a TW PDF at long times prior to saturation \cite{Prolhac2011}, and this reflects into nontrivial time evolution for ${\cal S}(t)$ and ${\cal K}(t)$.

\section{Dynamic renormalization group}\label{sec:analytic}

As a first exploration into the scaling behavior predicted by Eq.\ \eqref{eq:n}, in this Section we study the following generalized version of it,
\begin{align}\label{eq:ng}
   \partial_t \phi= A_n& \partial_x^2 \phi - B_n \partial_x^4 \phi + \lambda_n \partial_x (\phi \partial_x^n\phi) + \eta , \\
   \langle \eta(x,t) \eta(x',t'&) \rangle = D_n \, \delta(x-x') \delta(t-t') , \nonumber
\end{align}
where, in the context of our present work, the $A_n$ term is of technical importance and only for $n=1$ and $n=3$, as $\partial_x^2 \phi $ terms emerge in the renormalization procedure for those cases. Indeed, $A_n=0$ in Eq.\ \eqref{eq:n}. While Eq.\ \eqref{eq:ng} allows us to treat different values of $n$ in parallel, it is useful to keep track of the $n$ index in the parameters. Nevertheless, to simplify the notation we will suppress it whenever possible without ambiguity.

We extract the scaling behavior predicted by Eq.\ \eqref{eq:ng} through a standard DRG approach \cite{Forster1977}, which has been successfully applied in the $n=0$ case \cite{Rodriguez-Fernandez2019}. We first rewrite Eq.\ \eqref{eq:ng} in Fourier space as
\begin{eqnarray}\label{eq:Fourier}
     \widehat{\phi}(k,\omega)= G_0(k,\omega) \widehat{\eta}(k,\omega) + G_0(k,\omega) \lambda_n {\rm i} k \times \nonumber \\
    \int_{-\infty}^{\infty} \frac{d\Omega}{2\pi} \int_{-\Lambda}^{\Lambda} \frac{dq}{2\pi} \ \widehat{\phi}(q,\Omega) ({\rm i} q)^n \widehat{\phi}(k-q,\omega-\Omega),
\end{eqnarray}
with $G_0(k,\omega)= [ {\rm i}\omega + A_n k^2 + B_n k^4 ]^{-1}$. In Eq.\ \eqref{eq:Fourier}, a wide hat denoted space-time Fourier transform, $k$ is one-dimensional wave-vector, $\omega$ is time frequency, and $\rm{i}$ is the imaginary unit. We now take an arbitrary positive parameter $\ell$ and separate slow ($|k|<\Lambda e^{-\ell}$, denoted by superindex $<$) from fast ($\Lambda e^{-\ell}<|k|<\Lambda$, denoted by superindex $>$) Fourier modes, i.e.\
$ \widehat{\phi}= \widehat{\phi}^> +  \widehat{\phi}^< $ and $ \widehat{\eta}=\widehat{\eta}^> + \widehat{\eta}^< $, where
$
    \widehat{\phi}^>=\widehat{\eta}^>=0 $ if $ |k|<\Lambda e^{-\ell} $ and $
    \widehat{\phi}^<=\widehat{\eta}^<=0 $ for $ |k|>\Lambda e^{-\ell} $. We expand the fast modes $\widehat{\phi}^>$ into a perturbative series in $\lambda_n$ as
$
\widehat{\phi}^> = \widehat{\phi}^>_0 + \lambda_n \widehat{\phi}^>_1 + \lambda_n^2 \widehat{\phi}^>_2 + \mathcal{O}(\lambda_n^3).
$
Using this expansion, we can rewrite Eq.\ \eqref{eq:Fourier} for the slow modes as
\begin{eqnarray}
G^{-1}(k,\omega) \widehat{\phi}^<(k,\omega)= \widehat{\eta}^<(k,\omega) - \lambda_n {\rm i} k \times \nonumber \\ \int_{-\infty}^{\infty} \frac{d\Omega}{2\pi} \int_{-\Lambda}^{\Lambda} \frac{dq}{2\pi} \ 
[1+\Gamma_n] \times \label{eq:G} \\ (iq)^n\widehat{\phi}^<(q,\Omega) \widehat{\phi}^<(k-q,\omega-\Omega), \nonumber 
\end{eqnarray}
where $ G(k,\omega) = [ {\rm i}\omega + A_n k^2 + B_n k^4 + \Sigma_n ]^{-1} $. The noise variance is obtained from the computation of $\langle \widehat{\phi}^<(k,\omega) \widehat{\phi}^<(-k,-\omega) \rangle$, which leads to 
\begin{equation}
    \langle \widehat{\eta}^<(k,\omega) \widehat{\eta}^<(k',\omega') \rangle = (D_n +\Phi_n) \delta_{k+k'} \delta_{\omega+\omega'}.
    \label{eq:Noise}
\end{equation}
The functions $\Sigma_n $, $\Gamma_n$, and $\Phi_n$ appearing in Eqs.\ \eqref{eq:G} and \eqref{eq:Noise} can be evaluated diagrammatically \cite{Forster1977,Rodriguez-Fernandez2025}.
Up to one loop and for $n=1,2$, and 3, 
\begin{eqnarray}
    \Sigma_1 &=& \lambda^2 D_1 \int^{>} \frac{\pi}{8q^2(B_1q^2+A_1)} k^2 - \nonumber \\  
     & & \frac{\pi^2 (4B_1^2 q^4 + 7 A_1B_1 q^2 + A_1^2)}{64q^4(B_1q^2+A_1)^4} k^4 dq , \\
    \Sigma_2 &=& \lambda^2 D_2 \int^>-\frac{3\pi}{32B_2^2q^6} k^4 dq , \\
    \Sigma_3 &=& \lambda^2 D_3 \int^{>} \frac{3\pi q^2}{8(B_3q^2+A_3)^2} k^2 - \nonumber \\  
     & & \frac{\pi (10B_3^2 q^4 + 17 A_3B_3 q^2 + A_3^2)}{64(B_3q^2+A_3)^4} k^4 dq.
\end{eqnarray}
Remarkably, ${\cal O}(k^2)$ terms appear in both $\Sigma_1$ and $\Sigma_3$, with signs for their coefficients which lead to morphological instabilities in the renormalized interface equations. This will offer an analytical explanation for the dynamical instabilities found in the computational simulations of Eq.\ \eqref{eq:n} for $n=1$ and $n=3$, to be discussed in Sec.\ \ref{sec:numeric}.

The $\Gamma_n$ terms which contribute to vertex (nonlinearity) renormalization in Eq.\ \eqref{eq:G} take nonzero values at $\mathcal{O}(k^0)$ for $n=1,2$, and 3 (see details in \ref{app:1}). On the other hand, due to the conserved nature of the deterministic terms in Eq.\ \eqref{eq:ng}, the contribution from the noise renormalization, $\Phi_n$, is $\mathcal{O}(k^2)$, hence the coarse-grained variance of the noise remains unchanged for non-conserved noise  \cite{Grinstein1991,Grinstein1995}, i.e., $D_n^<=D_n$.

If we rescale space and time as $\tilde{x}=bx,\ \tilde{t}=b^z t$, and $ \tilde{\phi}=b^{-\alpha} \phi (\tilde{x},\tilde{t})$, Eq.\ \eqref{eq:ng} reads, after dropping the tildes,
\begin{align}
    \partial_t \phi = A_n b^{z-2} \partial_x^2 \phi &- B_n b^{z-4} \partial_x^4 \phi \nonumber \\
    & + \lambda_n b^{\alpha+z-(n+1)} \partial_x (\phi \partial_x^n \phi) + \eta , \\
    \langle \eta(x,t) \eta(x',t') \rangle & = 
    D_n b^{z-2\alpha-1} \, \delta(x-x') \delta(t-t') . \nonumber
    \label{eq:rescaled}
\end{align}
Finally, we can write down a set of RG flow equations for the renormalized (coarse-grained and rescaled) parameters $\tilde{A}_n=A_n^< b^{z-2}$, $\tilde{B}_n=B_n^< b^{z-4}$, $\tilde{\lambda}_n= \lambda_n b^{\alpha+z-n-1}$, and $\tilde{D}_n=D_n b^{z-2\alpha-1}$. 
By taking $b=e^{\ell}$, in the limit of vanishing $\ell$ 
these flow equations take the form
\begin{eqnarray}\label{eq:flujo}
    \frac{d \lambda_n}{d\ell} &=& \lambda_n(\alpha + z - n - 1+\mathcal{L}_n), \nonumber  \\
    \frac{d D_n}{d\ell} &=& D_n(z-2\alpha-1), \nonumber \\
    \frac{d A_n}{d\ell} &=& A_n (z-2 + \mathcal{A}_n ), \\
    \frac{d B_n}{d\ell} &=& B_n \left( z- 4 + \mathcal{B}_n \right), \nonumber 
\end{eqnarray}
where $\mathcal{L}_n$, $\mathcal{A}_n$, and $\mathcal{B}_n$ contain the nontrivial, $n$-dependent renormalization contributions to the nonlinear and linear parameters in Eq.\ \eqref{eq:ng}.

For any value of $n$, the RG flow equations, Eqs.\ \eqref{eq:flujo}, have a linear, Gaussian, or weak-coupling fixed point at which $\lambda_n=A_n=0$, such that $B_n, D_n \neq 0$. This corresponds to the linearized Eq.\ \eqref{eq:n}, usually termed the linear MBE equation \cite{Barabasi95,Krug1997}, whose scaling behavior is characterized by the $n$-independent exponent values $\alpha_{\rm G}=3/2$ and $z_{\rm G}=4$, so that $\beta_{\rm G}=3/8$. 

In addition to the Gaussian fixed point, the RG flow, Eqs.\ \eqref{eq:flujo}, can in principle feature non-linear or strong coupling fixed points. The corresponding scaling exponent values will depend on $n$ and must fulfill the following scaling relation which guarantees that $D_n\neq 0$ at the fixed point,
\begin{equation}
    2\alpha+d = z  , \; \mathrm{(Hyperscaling)} \label{eq:hyp}
\end{equation}
with $d=1$. This scaling relation ---termed hyperscaling due to its $d$-dependent form \cite{Barabasi95,Krug1997}---, is expected to hold to all orders in perturbation theory as it expresses the fact that a conserved nonlinearity cannot renormalize non-conserved noise \cite{Grinstein1991,Grinstein1995}. 

If $\Gamma_n$ is identically zero or negligible at $\mathcal{O}(k^0)$ ---so that $\mathcal{L}_n=0$ in Eq.\ \eqref{eq:flujo}---, the exponents at the nonlinear fixed point, $\lambda_n\neq0$, are such that
\begin{equation}
    \alpha+z = n+1 . \; \mathrm{(Galilean)} \label{eq:Gal}
\end{equation}
For $n=0$, Eq.\ \eqref{eq:Gal} holds exactly at the nonlinear fixed point, reflecting the Galilean symmetry of the Burgers equation \cite{Forster1977,Rodriguez-Fernandez2019}, hence the name of this scaling relation. However, strictly speaking Eq.\ \eqref{eq:n} is not Galilean invariant for $n>0$, in which case the scaling relation, Eq.\ \eqref{eq:Gal} is a generalization of the Galilean case which need not necessarily hold at the strong coupling fixed point.

For later convenience, in Table \ref{tab:resumen} we collect the values of the scaling exponents obtained in Sec.\ \ref{sec:numeric} from numerical simulation for $n=1,2$, and $3$, together with the values (in square brackets) obtained for each $n$ by \emph{simultaneously} imposing the hyperscaling and the Galilean scaling relations, Eqs.\ \eqref{eq:hyp} and \eqref{eq:Gal}.
\begin{table}
    \centering
    \begin{tabular}{|c||c|c|c|}
    \hline
    $n$ & $\alpha_{\rm num}$ & $z_{\rm num}$ & $\beta_{\rm num}$ \\
    \hline \hline
       0 & 0 [0] & 1 [1] & 0 [0] \\
       \hline
       1  & 0.45 [1/3] & 1.88 [5/3] & 0.24 [1/5] \\
       \hline
       2  & 0.67 [2/3] & 2.3 [7/3] & 0.29 [2/7] \\
       \hline
       3  &  1.4 [1] & 3.8 [3] & 0.37 [1/3] \\
    \hline
    \end{tabular}
    \caption{Summary of the approximate values of the strong coupling critical exponents obtained numerically in Sec.\ \ref{sec:numeric} for $n=1,2$, and 3. Values in brackets are the one-loop DRG predictions derived in Sec.\ \ref{sec:analytic} considering $\Gamma_n\simeq0$ [hence $\mathcal{L}_n \simeq 0$ in Eq.\ \eqref{eq:flujo}]. For completeness, the values for $n=0$ are taken from Ref.\ \cite{Rodriguez-Fernandez2019}.}
    \label{tab:resumen}
\end{table}
Two particular cases are noteworthy. First, for $n=0$ Eqs.\  \eqref{eq:hyp} and \eqref{eq:Gal} retrieve the analytical \cite{Forster1977} and numerical \cite{Rodriguez-Fernandez2019} exponent values ($\alpha_0=0$, $z_0=1$) of the stochastic Burgers equation. This is because both the hyperscaling and Galilean relations hold (and are believed to be exact \cite{Forster1977}) for this system. On the other hand, for $n=2$, the numerical values obtained for the scaling exponents are almost equal to those that result from the set of two linear equations for two unknowns posed by requesting that Eqs.\ \eqref{eq:hyp} and \eqref{eq:Gal} simultaneously hold. In contrast, for $n=1$ and $n=3$ the numerical exponents obtained numerically in Sec.\ \ref{sec:numeric} fulfill the hyperscaling relation, but seem far from fulfilling the Galilean relation, suggesting that $\Gamma_n\neq0$ (while $\Phi_n=0$) for these values of $n$.


\section{Numerical simulations}\label{sec:numeric}

The DRG analysis just presented allows for the possibility of nonlinear or strong coupling behavior in the time evolution predicted by Eq.\ \eqref{eq:n}. However, the analysis is perturbative in nature and alternative confirmation of this result is required, especially in view of the limitations of one-loop predictions for related systems like the cKPZ equation \cite{Janssen1997}. In this section we resort to numerical simulations of Eq.\ \eqref{eq:n} for the cases which are most frequently found in the literature, $n=1,2$, and 3.

For each value of $n$, we have performed numerical simulations of Eq.\ \eqref{eq:n} using the most suitable scheme from the point of view of numerical stability, as specified in Appendix \ref{app:3}. The schemes employed for $n=1$ and 3 are based on finite differences ---implementing and generalizing, respectively, the proposal made in Ref.\ \cite{Sasamoto09} for the Burgers nonlinearity--- while the scheme employed for $n=2$ is a pseudospectral method. In all cases, periodic boundary conditions have been used. Further details on the numerical methods are presented in \ref{app:3}.

%
%


\subsection{Deterministic equation: $x\leftrightarrow-x$ vs $n$}\label{sec:deterministic:sinus}

We begin by performing a numerical integration of the deterministic case ($D_n=0$) of Eq.\ \eqref{eq:n} because this can provide us with intuition on the behavior of the equation as a function of the value of $n$. For these numerical results, we impose a sinusoidal initial condition $\phi(x,t=0)=\sin(k_0x)$ for a fixed value of $k_0$. This function is chosen because of its simplicity, generality, and up-down symmetry. Then Eq.\ \eqref{eq:n} is solved numerically for $n=0,1,2,$ and $3$.

\begin{figure}[t!]
    \centering
\includegraphics[width=0.5\textwidth]{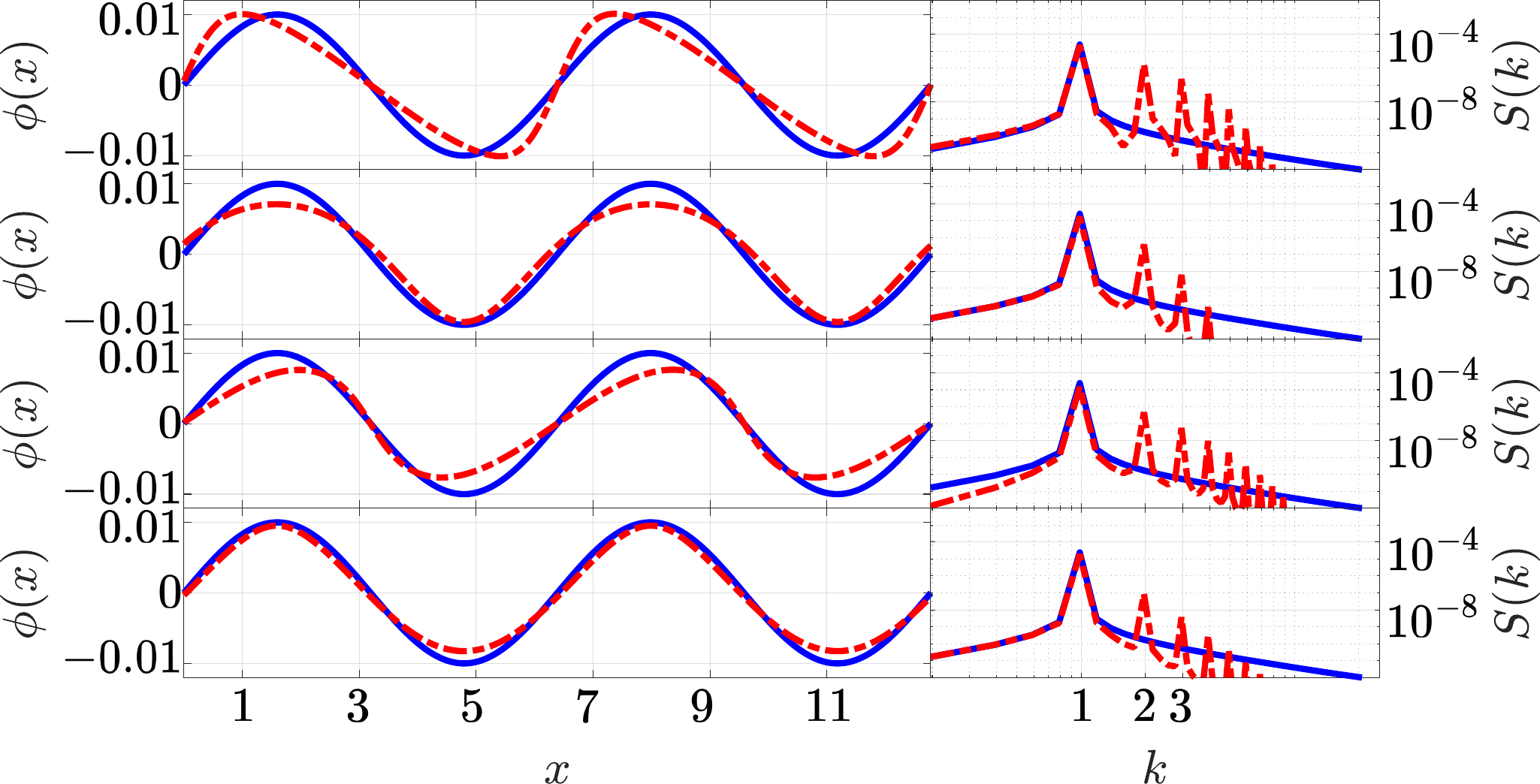}
\caption{Deterministic ($D_n=0$) solutions of Eq.\ \eqref{eq:n} (left panels) and their PSDs (right panels), for $n=0,1,2,$ and 3, top to bottom. In each panel, the solid blue line corresponds to the initial condition and the dash-dotted red line corresponds to the numerical solution at time $t_f$. Top to bottom, $n=0$, $\lambda_0=2^6$, $B_0=2^{-6}$, $t_f=2^{-1}$ (first row); $n=1$, $\lambda_1=2^6$, $B_1=2^{0}$, $t_f=2^0$ (second row); $n=2$, $\lambda_2=2^8$, $B_2=2^{-2}$, $t_f=2^{-2}$ (third row); $n=3$, $\lambda=2^8$, $B_3=2^{1.4}$, $t_f=2^{-4}$ (fourth row). In all cases, the initial condition is $\phi(x,t=0)=A\sin(k_0 x)$ with $A=0.01$ and $k_0=1$.}
    \label{fig:detPsin-n1234}
\end{figure}

\begin{figure}[t!]
    \centering
\includegraphics[width=0.5\textwidth]{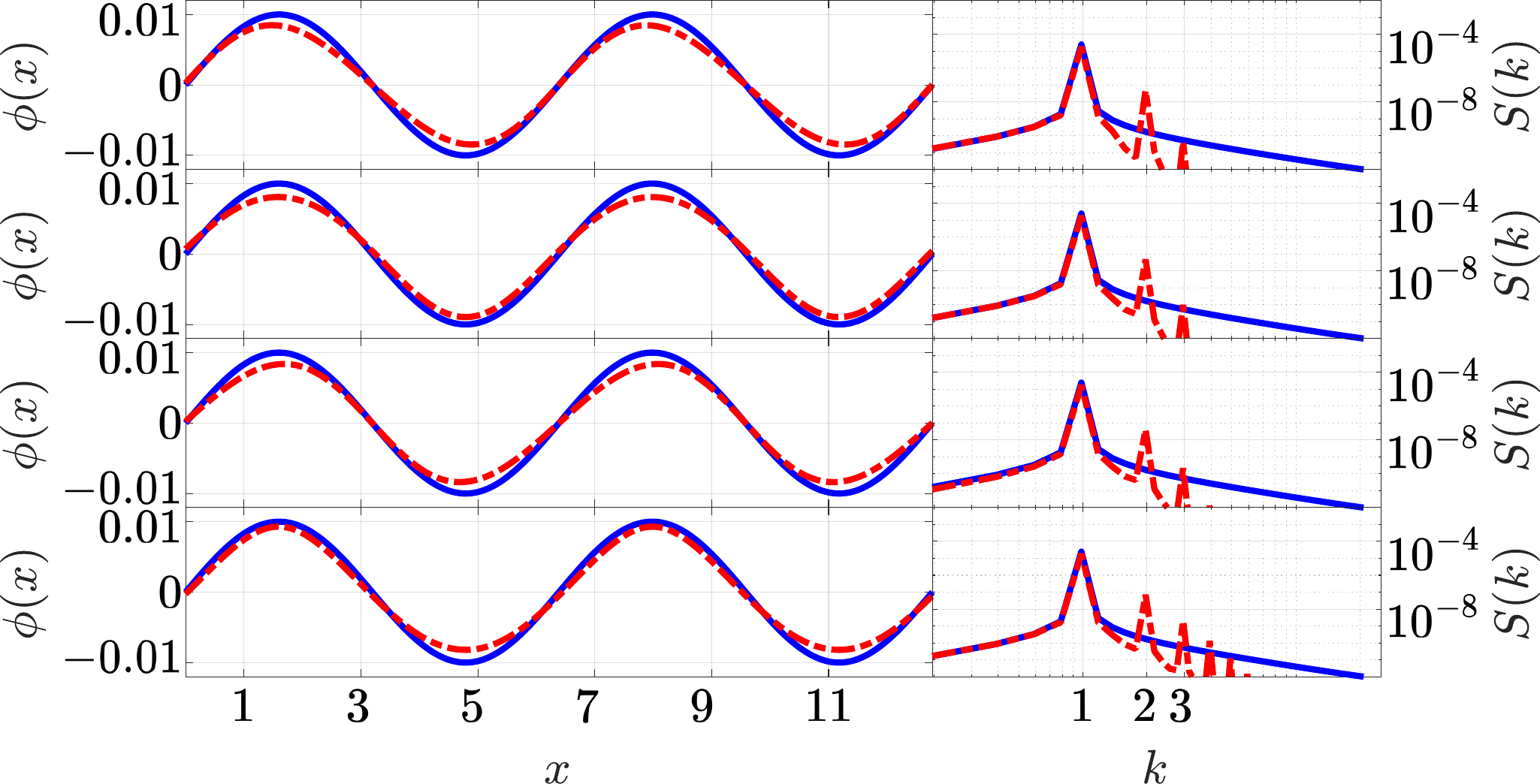}
\caption{Same as Fig.\ \ref{fig:detPsin-n1234} but for $n$-independent simulation parameter values $\lambda_n=2^8$, $B_n=2^{-\frac{1}{2}}$, and $t_f=2^{-2}$. Results are for $n=0$, $n=1$, $n=2$, and $n=3$, top to bottom. The initial condition is again $\phi(x,t=0)=A\sin(k_0 x)$ with $A=0.01$ and $k_0=1$.}
    \label{fig:detsin-n1234}
\end{figure}

Different parameters have intentionally been used for each value of $n$ to better underscore the main features of each case. Results are displayed in Fig.\ \ref{fig:detPsin-n1234},
which contains the numerical solutions for $\phi(x,t)$, together with their structure factors, $S(k,t)$, at long times. In all cases the profile amplitude decreases with increasing time, but different values of $n$ differ in the way in which the front profile is distorted. For $n=0$ and 2, it is sheared along the $x$ direction, with peaks and valleys being displaced in opposite directions and producing a shape which is reminiscent of the saw-tooth structures from Burgers equation \cite{Burgers1974,Bec2007,Bendaas2018}. This is probably not surprising, as the $n=0$ equation features the Burgers' non-linearity. For odd $n=1$ and 3 values, the locations of local extrema are not shifted, but the peaks of the front profile decrease at a different rate than the valleys. For $n=1$ the peaks are reduced faster whereas valleys are for $n=3$. On the other hand, the up-down symmetry of the initial condition is lost for all values of $n$, as indicated by the emergence of even harmonics in the structure factors. The amplitudes of the harmonics increase with increasing $n$ as suggested by Fig.\ \ref{fig:detsin-n1234}), where results of simulations are shown in which $\lambda_n$ and $B_n$ are given 
$n$-independent values. This means that energy transfer from the external stochastic driving to smaller scales is more efficient for larger $n$.

The lack of up-down symmetry is explicit in Eq.\ \eqref{eq:n}, and becomes more evident after expanding the non linear term as
\begin{equation}\label{sec:deterministic:eq:nlnum}
\partial_x(\phi \partial_{x}^n\phi)=(\partial_x\phi)(\partial_{x}^n\phi)+\phi\partial_{x}^{n+1}\phi .
\end{equation}
Indeed, the second term on the right-hand side depends explicitly on $\phi$ itself, affecting unevenly the top and bottom regions of the $\phi(x)$ profile. This modulation is all the more significant when $\phi$ changes from positive to negative values.

The interpretation just derived from Fig.\ \ref{fig:detPsin-n1234} is supported by the results shown in Fig.\ \ref{fig:detsin-n1234} for $n$-independent parameter values. Namely, regardless of $n$ the amplitude of an initial sinusoidal condition damps out and, while for $n=0$ and 2 the deterministic terms in Eq.\ \eqref{eq:n} produce saw-tooth shapes, for $n=1$ and 3 the symmetry of the initial condition under space inversion $x\leftrightarrow-x$ is preserved along the time evolution. Mathematically, this behavior manifests explicitly the fact that odd (even) $n$ makes Eq.\ \eqref{eq:n} invariant (non-invariant) under space inversion. Physically, it agrees qualitatively with the role of even-$n$ nonlinearities as models of convective transport and of odd-$n$ nonlinearities as descriptions of smoothening by capillarity-like effects; recall the various physical contexts in which these nonlinearities can be found as discussed in Sec.\ \ref{sec:intro}.


\subsection{Stochastic equation: $n=1$ }

We next proceed with our numerical simulations of the full stochastic Eq.\ \eqref{eq:n} for several representative values of $n$. First we address the $n=1$ case, which reads
\begin{equation}\label{eq:n=1}
   \partial_t \phi=- B_1 \partial_x^4 \phi + \lambda_1 \partial_x (\phi \partial_x\phi) + \eta , 
\end{equation}
where the nonlinear term is e.g.\ a particular case of the so-called porous medium equation, see Sec.\ \ref{sec:intro}. For our numerical study, we have employed a finite-difference scheme detailed in \ref{app:3}. As further discussed in \ref{app:2}, the numerical implementation of the equation is susceptible of becoming unstable whenever $\phi<0$. To avoid this difficulty, the numerical integration has been performed starting out from an initial condition placed at a constant value $\phi_0=\phi(t=0)>0$, checking that the solution remains positive for all $x$ and $t$.

\begin{figure}[t!]
    \centering
\includegraphics[width=0.5\textwidth]{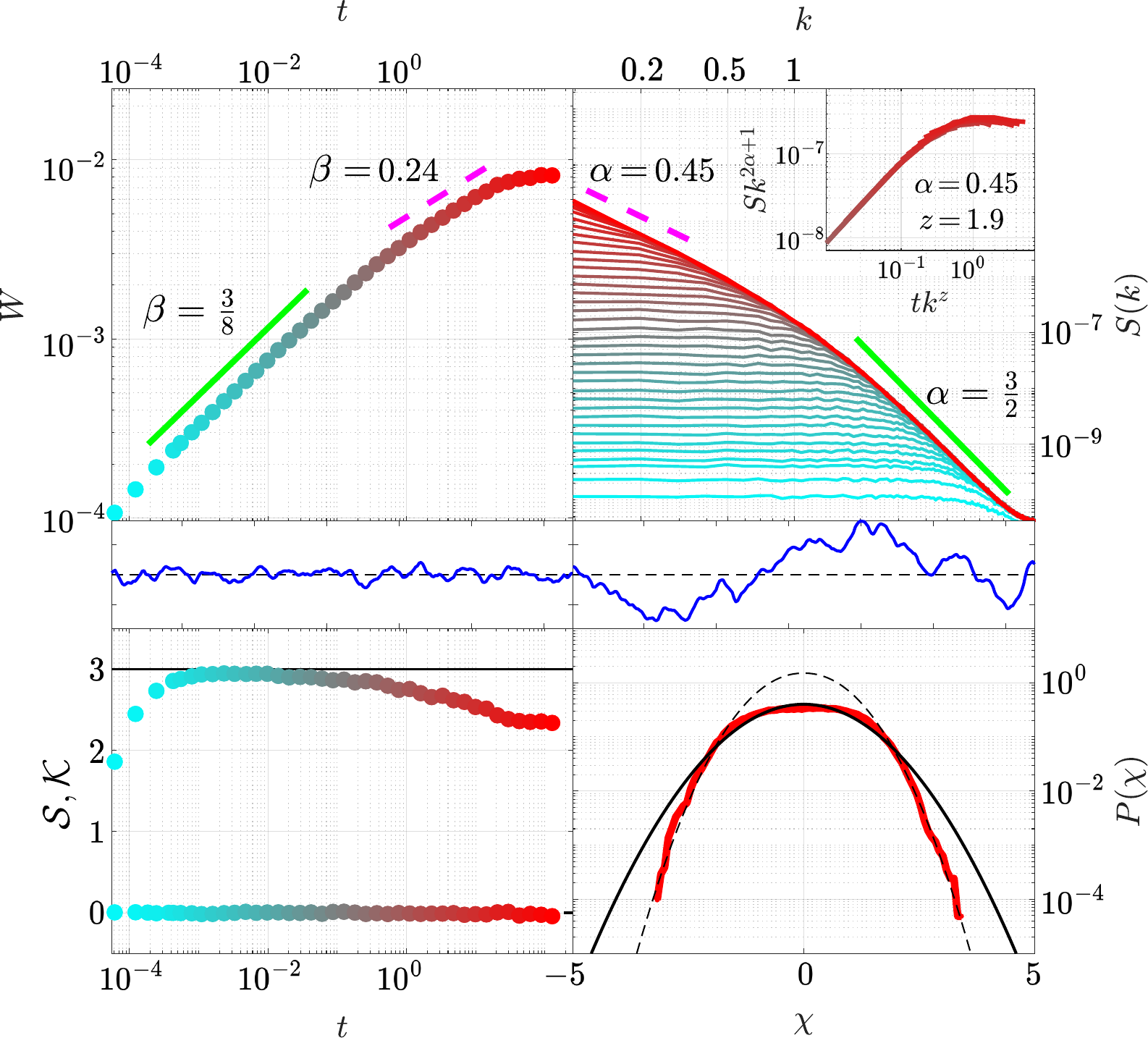}
\caption{Results for numerical simulation of Eq.\ \eqref{eq:n} for $n=1$, namely, Eq.\ \eqref{eq:n=1}. Parameters are: $L=64$, $T_t=2^8$, $B_1=2^0$, $\lambda_1=2^2$, $D_1=2^{-14}$, and
$\phi_0=2^{-2}$. Averages are over 300 realizations of the noise. Top left, roughness $W(t)$ over time. Top right, PSD at different times. Top right inset, collapse of the PSD data. Bottom left, skewness (lower symbols) and kurtosis (upper symbols) for different times. Bottom right, histogram of height fluctuations at saturation to steady state. Middle left, morphology during linear growth. Middle right, morphology at saturation. Measurements are taken at times $t=\Delta t \cdot 2^N$ with $N=0,1,2,3,\ldots$ Blue (red) dots and lines correspond to early (late) times. Dashed purple and solid green reference straight lines on the top panels have the slopes indicated by the indicated scaling exponent values. Solid and dashed lines in the bottom right panel correspond to two different Gaussian distributions.}
    \label{fig:N1-Sto}
\end{figure}

The middle panels in Fig.\ \ref{fig:N1-Sto} show two examples of morphologies for $\phi(x,t)$, for relatively short (left middle panel) and long times (at saturation, right middle panel), highlighting substantial changes along the time evolution. More quantitatively, Fig.\ \ref{fig:N1-Sto} top left shows the evolution of the roughness $W$ over time, which is consistent with power-law behavior as $W \sim t^\beta$ prior to saturation to steady state at the longest times. 
Before those, the growth exponent $\beta$ is quite close to the linear MBE value $\beta\simeq 3/8$ for short times; for longer times when the non-linearity in Eq.\ \eqref{eq:n=1} becomes relevant, the growth exponent crosses over to a different value, $\beta\approx0.24$. We can estimate the value of the roughness exponent within each regime from the PSD of $\phi$. Indeed, within each one of the two growth regimes just identified, $S(k,t)$ displays dynamic scaling as described by Eq.\ \eqref{eq:psd_fv}, with $\alpha$ being close to $3/2$ as for the linear MBE equation dominating at high wave vector values and a strong coupling value $\alpha\approx0.45$ at low $k$; see the top right panel in Fig.\ \ref{fig:N1-Sto}. We assess this scaling behavior for the nonlinear regime in further detail on the top right inset of Fig.\ \ref{fig:N1-Sto}, which displays a data collapse of the PSD curves for large length scales and long times, according to Eq.\ \eqref{eq:psd_fv}. Specifically, rescaled large-scale data for $S(k,t)$ are shown with $t\in [2\cdot 10^{-1}, 2\cdot 10^1]$ and $k\in (0, 0.5]$. A consistent collapse is indeed achieved for $\alpha=0.45$ and $z=1.9$, which in turn yields $\beta =\alpha/z\approx0.24$, consistent with the growth exponent value estimated for the roughness $W(t)$ at long times. Here and in what follows, the uncertainty in our computed exponent values is one unit in the last quoted digit.

The bottom panels in Fig.\ \ref{fig:N1-Sto} asses the fluctuation statistics of $\phi$ for Eq.\ \eqref{eq:n=1}. The measured skewness remains quite close to zero for all times. However, the kurtosis changes from a Gaussian value during the linear regime, to a substantially smaller value that becomes time-independent at saturation. The full normalized PDF of $\phi(x,t)$ fluctuations at saturation to steady state, shown in the bottom right panel, agrees with these measurements, being symmetric but non-Gaussian. Actually, the tails of this histogram are not far from Gaussian decay, but the central region remains closer to uniform. Overall, these data suggest a time crossover from a Gaussian distribution governing field fluctuations at short times (as for the linear MBE equation) to non-linear behavior at longer times and at saturation. In this regime, an \textrm{emergent symmetry} seems to be occurring since the statistics are symmetric (zero value for ${\cal S}$) while the nonlinear equation is not up-down ($\phi \leftrightarrow -\phi$) symmetric, see Sec.\ \ref{sec:discussion}.

\subsection{Stochastic equation: $n=2$}

We next perform a similar numerical study of Eq.\ \eqref{eq:n} for $n=2$, namely for the equation
\begin{equation}\label{eq:n=2}
   \partial_t \phi=- B_2 \partial_x^4 \phi + \lambda_2 \partial_x (\phi \partial_x^2\phi) + \eta .
\end{equation}
The nonlinear term $\phi\partial_x^2 \phi$ in this continuum model occurs crucially in the Sawada-Kotera or Caudrey-Dodd-Gibbon, soliton bearing equation \cite{Sawada1974,Caudrey1976,Zwillinger2021}. As a current, it has been recently obtained e.g.\ for diffusion of particles subject to additional constraints beyond mass conservation \cite{Han2024}.

\begin{figure}[b!]
    \centering
\includegraphics[width=0.5\textwidth]{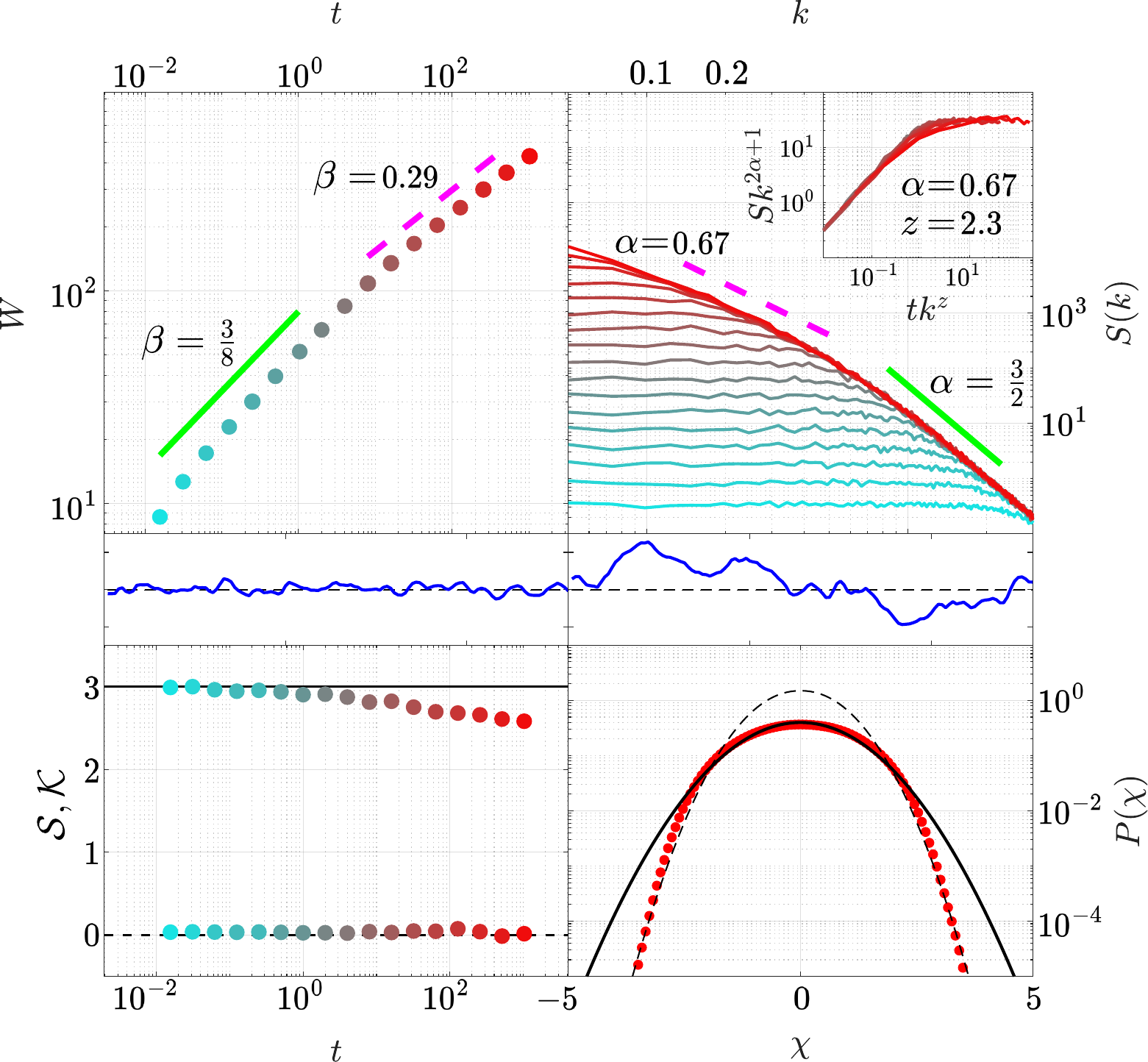}
\caption{Results for numerical simulation of Eq.\ \eqref{eq:n} for $n=2$, namely, Eq.\ \eqref{eq:n=2}. Parameters: 256 elements of size $\Delta x=2^{0}$, $2^{20}$ time steps of duration $\Delta t=2^{-9}$, $B_2=\lambda_2=D_2=1$,
and $\phi_0=0$. Averages are over 200 realizations of the noise. The descriptions for all the panels are analogous to those of Fig.\ \ref{fig:N1-Sto}.}
    \label{fig:N2-Sto}
\end{figure}

We simulate Eq.\ \eqref{eq:n=2} by means of a pseudospectral scheme, as detailed in \ref{app:3}. Sample morphologies for $\phi(x,t)$ are shown in the middle panels of Fig.\ \ref{fig:N2-Sto} for short (left middle panel) and long times at saturation (right middle panel).
The remaining panels of the same figure are also completely analogous to those of Fig.\ \ref{fig:N1-Sto} but for the case at hand and are qualitatively similar in the sense of describing crossover behavior between a short time regime dominated by the linear MBE behavior to a novel 
strong-coupling regime quite close to that predicted for Eq.\ \eqref{eq:n=2} in Sec.\ \ref{sec:analytic}, recall Table \ref{tab:resumen}. Thus, the power-law behavior of the roughness in the top left panel of Fig.\ \ref{fig:N2-Sto} crosses over in time from $\beta\simeq 3/8$ at short times to $\beta\simeq 2/7$ at long times prior to saturation. The collapse of large-scale $S(k,t)$ data in the top right panel of Fig.\ \ref{fig:N2-Sto} is consistent with the strong coupling exponents $\alpha=2/3$ and $z=7/3$ predicted from the hyperscaling and Galilean scaling relations, Eqs.\ \eqref{eq:G} and \eqref{eq:hyp} ---that in turn imply $\beta=2/7$, consistent with the $W(t)$ data---, while the roughness exponent at short distances (large $k$ values) is close to the $3/2$, linear MBE value. The time evolution of the field statistics is also consistent with this overall behavior of crossover between linear and nonlinear scaling behavior while keeping a symmetirc distribution for all times. Thus, the skewness again fluctuates closely around zero regardless of the time regime, with the kurtosis shifting from 3 down to a smaller value for long times. At saturation, the field PDF remains symmetric and, while not agreeing with a simple Gaussian form, displays Gaussian decay at its tails nonetheless.


\subsection{Stochastic equation: $n=3$ }

We finally address the $n=3$ case of Eq.\ \eqref{eq:n} through numerical simulations, namely,
\begin{equation}\label{eq:n=3}
   \partial_t \phi=- B_3 \partial_x^4 \phi + \lambda_3 \partial_x (\phi \partial_x^3\phi) + \eta .
\end{equation}
The nonlinear term in this equation is a well-known weakly-nonlinear description of relaxation via capillarity or surface tension effects in thin fluid films, cf.\ Sec.\ \ref{sec:intro}, 
and recently appears in competition with other nonlinearities in the so-called active model B+, relevant to active matter \cite{Tjhung2018,Caballero2018b}. 

Equation \eqref{eq:n=3} is more demanding from the numerical point of view than its $n=1$ and 2 counterparts. We have simulated it numerically by means of a finite-difference scheme that adapts to the present case the scheme originally proposed in Ref.\ \cite{Sasamoto09} for the stochastic Burgers equation, see  \ref{app:3}. Akin to the $n=1$ case, Eq.\ \eqref{eq:n=1}, numerical stability issues also require us to start out from a positive initial condition, checking that this property is preserved by the time evolution, see \ref{app:2}. 

\begin{figure}[t!]
    \centering
\includegraphics[width=0.5\textwidth]{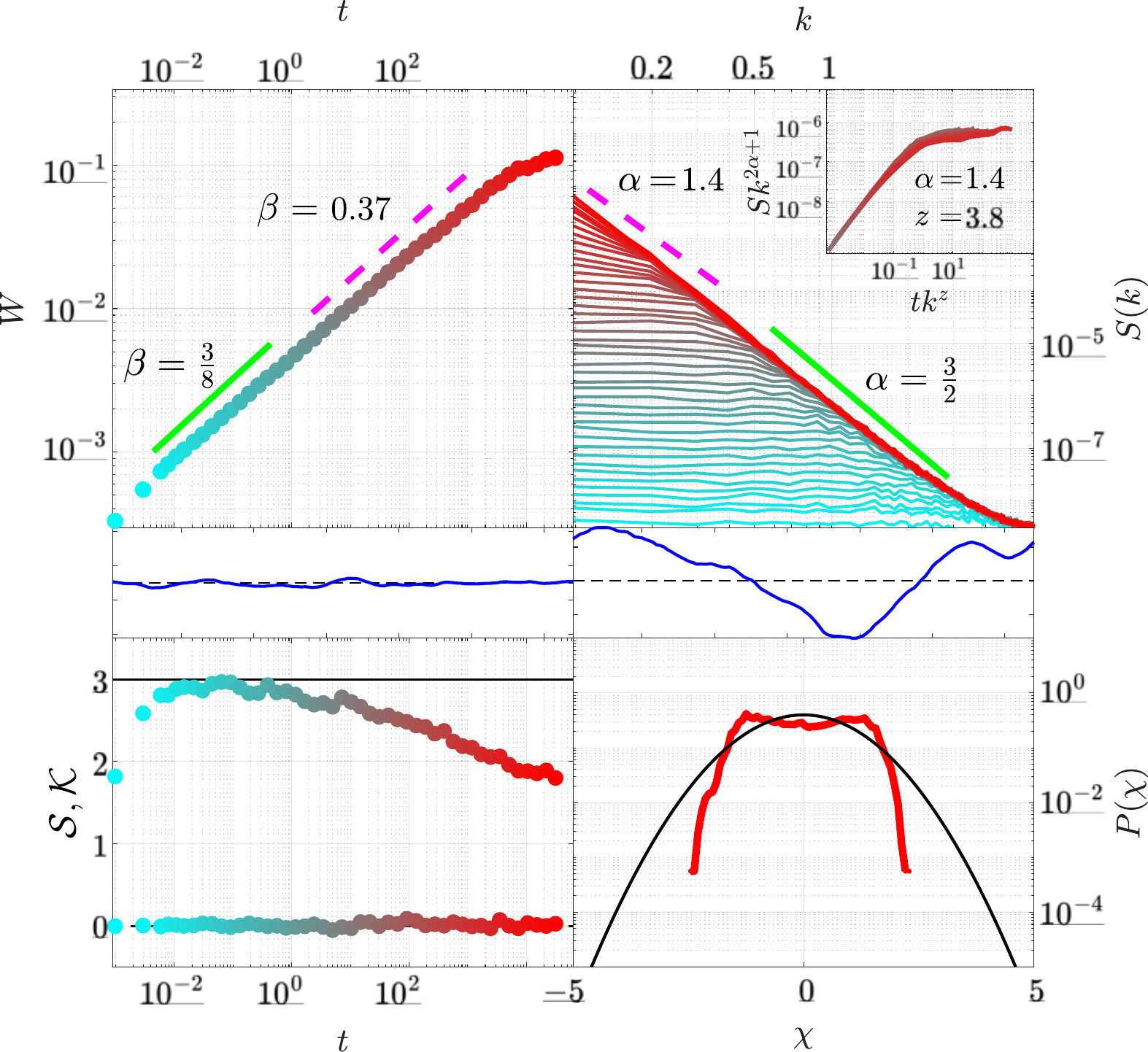}
\caption{Results for numerical simulation of Eq.\ \eqref{eq:n} for $n=3$, namely, Eq.\ \eqref{eq:n=3}. Parameters: 256 elements of size $\Delta x=2^{-1}$, $2^{27}$ time steps of duration $\Delta t=2^{-10}$, $B_3=2^{-2}$, $\lambda_3=2^2$, $D_3=2^{-14}$, and
and $\phi_0=2^{-1}$. Averages are over 700 realizations of the noise. The descriptions for all the panels are analogous to those of Figs.\ \ref{fig:N1-Sto}.}
    \label{fig:N3-Sto}
\end{figure}

The middle panels of Fig.\ \ref{fig:N3-Sto} provide sample morphologies of $\phi(x,t)$ according to Eq.\ \eqref{eq:n=3} for short (left middle panel) and long times at saturation (right middle panel).
Amplitude evolution is more substantial than for the previous values of $n$ studied in this section, being visually reminiscent of that obtained for the linear MBE equation \cite{Barabasi95,Krug1997}. Note, this arises as a nontrivial interplay between the deterministic and the stochastic terms in the equation and does not contradict the smoothening role of the former as assessed in Sec.\ \ref{sec:deterministic:sinus}. Quantitatively, the fact that the two deterministic terms are of the same (fourth) order in space derivatives for Eq.\ \eqref{eq:n=3} (in contrast with $n=0,1$, or 2) seems to somewhat complicate the identification of the scaling exponents. Thus, the time evolution of the roughness according to Eq.\ \eqref{eq:n=3} is shown on the top left panel of Fig.\ \ref{fig:N3-Sto}. The $\beta_{\rm G}=3/8$ growth exponent value corresponding to the linear MBE equation seems to hold for very short times only, while longer times are well accounted for by a very similar value $\beta\approx 0.37$. Likewise, the evolution of the structure factor $S(k,t)$ displayed in the top right panel of the same figure suggests that, while the short distance behavior agrees with the $\alpha_{\rm G}=3/2$ roughness exponent of the linear MBE equation, large scale properties are accounted for by exponent values (see the data collapse in the inset) $\alpha\approx 1.4$ and $z\approx 3.8$ (consistent with $\beta\approx 0.37$ as measured for $W(t)$ at long times) which are also quite close to the values for the linear equation (recall $z_{\rm G}=4$). Still, we believe that this large-scale behavior is dominated by nonlinear effects, given that the fluctuation statistics are not Gaussian, see the bottom panels of Fig.\ \ref{fig:N3-Sto}. Thus, although 
the skewness remains close to zero for all times, the kurtosis drastically departs from its Gaussian value in the nonlinear regime at intermediate and long times. Furthermore, the field PDF at the longest times, while symmetric, no longer resembles a Gaussian distribution, instead a bimodal shape seems to occur which differs quite dramatically from the simple Gaussian behavior that holds for the linear MBE equation.



\section{Discussion}\label{sec:discussion}


Both for $n=0$ and 2, the values of the scaling exponents that we find in our numerical simulations of Eq.\ \eqref{eq:n} agree quite well with those theoretically predicted on the basis of simultaneous hyperscaling and Galilean scaling relations; see Table \ref{tab:resumen2}, where we assess the degree to which each of these scaling relations is satisfied by the numerical values of the strong coupling exponents obtained in Sec.\ \ref{sec:numeric} for each value of $n$. The smaller each entry in the table is, the better the scaling relation of that column is numerically verified. Note that hyperscaling is quite approximately satisfied in all cases. Actually, for the reasons stated above this scaling relation is expected to hold exactly, so that deviations from it can be possibly reduced by employing larger system sizes and larger statistics in the numerical simulations. In contrast, odd values of $n=1, 3$, do not seem to agree with the Galilean scaling relation, agreement seemingly deteriorating with increasing $n$. 
\begin{table}
    \centering
    \begin{tabular}{|c||c|c|}
    \hline
    $n$ & $\begin{array}{c}
    \mathrm{Hyperscaling\!:} \\ |2\alpha_{\rm num}+1 -z_{\rm num}|
    \end{array}$
     & $\begin{array}{c}
    \mathrm{Galilean\!:} \\ |\alpha_{\rm num}+z_{\rm num}-(n+1)|
    \end{array}$ \\
    \hline \hline
       0 & 0 & 0 \\
       \hline
       1 & 0.02 & 0.33 \\
       \hline
       2 & 0.04 & 0.03 \\
       \hline
       3 & 0 & 1.2 \\
    \hline
    \end{tabular}
    \caption{Approximations to the hyperscaling (second column) and Galilean (third column) scaling relations, Eqs.\ \eqref{eq:hyp} and \eqref{eq:Gal}, according to the numerical results on the strong coupling critical exponent values obtained in Sec.\ \ref{sec:numeric} for $n=1,2$, and 3 and collected in Table \ref{tab:resumen}. For the second and third columns, the smaller the quoted value, the better the corresponding scaling relation is satisfied. For completeness, the values for $n=0$ are taken from Ref.\ \cite{Rodriguez-Fernandez2019}.}
    \label{tab:resumen2}
\end{table}
This behavior seems akin to known results for the 1D cKPZ equation, for which one-loop DRG \cite{Barabasi95} and a self-consistent expansion \cite{Katzav2002} both predict a so-called Galilean relation of the form $\alpha_{\rm cKPZ}+z_{\rm cKPZ}=4$ (which coincides with our $n=3$ Galilean relation, motivating our  name choice in the first place), while two-loop DRG \cite{Janssen1997} and detailed numerical simulations \cite{Carrasco2016} are consistent with failure of such a scaling relation. Hence, one might expect higher-order (e.g.\ two-loop) DRG analysis of Eq.\ \eqref{eq:n} to yield $n$-dependent corrections which might (not) cancel for even (odd) values of $n$. Note, the exponents for the $n=3$ case of Eq.\ \eqref{eq:n} seem even further off Galilean scaling than those of the cKPZ equation.


With respect to the statistics of height fluctuations, our numerical results seem consistent with zero skweness (as for the Gaussian distribution) but non-Gaussian kurtosis values irrespective of $n$, provided $n>0$ (recall that, in spite of displaying non-linear exponent values, the height statistics is Gaussian for the $n=0$ Burgers equation with non-conserved noise \cite{Rodriguez-Fernandez2019}). For $n=2$, the symmetry of the height PDF can perhaps be related with the system invariance under a combined $(x \leftrightarrow -x, \phi \leftrightarrow -\phi)$ transformation, as is the case for $n=0$ \cite{Rodriguez-Fernandez2019}. However, such a symmetry does not occur for odd $n$ values, for which the zero-skewness value can then be interpreted as a(n emergent) symmetry of the strong coupling fixed point which is broken at the level of the bare or unrenormalized height equation \cite{Batista2002}.

It would be interesting to consider if the strong coupling behavior found for Eq.\ \eqref{eq:n} generalizes to higher space dimensions. Indeed, the most relevant case for many relaxational dynamics models of recent interest \cite{Tjhung2018,Caballero2018,Caballero2018b,Caballero2020}, including active models A and B+, corresponds to $d=2$. In principle, generalizing Eq.\ \eqref{eq:n} to $d$ dimensions requires separately considering (this difficulty is analogous to that found in the generalization of the stochastic Burgers equation to $d>1$ \cite{Vivo2012,Vivo2014,Rodriguez-Fernandez2019,Rodriguez-Fernandez2020}) the case of odd $n=2p+1$, for which one would write
\begin{equation}\label{eq:2p+1}
   \partial_t \phi=- B_{2p+1} \nabla^4 \phi + \lambda_{2p+1} \nabla \cdot [\phi \nabla (\nabla^{2p}\phi)] + \eta(\mathbf{r},t) ,
\end{equation}
and the case of even $n=2p$, for which one would write
\begin{equation}\label{eq:2p}
   \partial_t \phi=- B_{2p} \nabla^4 \phi + \lambda_{2p} \sum_{j=1}^d \partial_{x_j} (\phi \partial_{x_j}^{2p} \phi) + \eta(\mathbf{r},t).
\end{equation}
Equation \eqref{eq:2p} actually generalizes for $n\neq0$ the continuum model of driven-diffusive systems of Ref.\ \cite{VanBeijeren1985}. 
For $p=0$ and $d=2$, numerical simulations and DRG analysis conclude that the fluctuation PDF is Gaussian. For $n>0$ and $d\geq 2$, this has been also been proven to be the case only very recently \cite{Cannizzaro2024}.
 


\section{Summary and Conclusions}\label{sec:conclusions}

We have introduced and studied a one-parameter family of stochastic equations indexed by a non-negative integer $n$, that include and generalize the stochastic Burgers equation ($n=0$ case), with the goal to further study conserved critical dynamics subject to non-conserved noise. The $n=3$ member of the family happens to be analogous to another celebrated continuum model of surface kinetic roughening, namely, the cKPZ equation. The nonlinearities of the various members of the family that we have addressed explicitly (for $n=1, 2$, and 3) appear in the continuum descriptions of various hard and soft condensed matter systems.

Working for the case of 1D interfaces, our main objective is to assess strong coupling behavior with respect to the scaling exponent values and also in terms of the statistics of height fluctuations. We have first approached the system behavior analytically through a one-loop DRG study, leading us to expect a hyperscaling relation to be satisfied irrespective of the value of $n$. In contrast, a Galilean-like scaling relation involving the critical exponents may or not be satisfied depending on vertex renormalization behavior, known in the cKPZ case to be possibly contingent upon the degree of the loop approximation made. Therefore, we have resorted to direct numerical simulations of the $n=1, 2$, and 3 systems. These results suggest that the vertex may not renormalize for even $n$, while we expect the converse to be the case for odd $n$ values, with vertex corrections possibly increasing with the value of $n$. The conclusion on the strong-coupling nature of the exponent values we assess is confirmed by the non-Gaussian behavior of the fluctuation PDF at saturation. For all $n$ values studied, and in spite of being symmetric (i.e., with zero skewness), the PDF seems to feature non-Gaussian kurtosis, being flatter than Gaussian in its central part with Gaussian-like tails. For even values of $n$, the behavior of the PDF is possibly related with the system symmetry under the combined $(x,\phi) \leftrightarrow (-x,-\phi)$ transformation. In contrast, systems with odd values of $n$ do not remain invariant under this transformation. Hence, for them the zero skewness of the fluctuation PDF can be interpreted as symmetry that, being broken at the level of the bare or unrenormalized evolution equations, emerges at their strong coupling fixed points that govern their asymptotic behaviors in time and space.

Technically, our numerical work with the odd-$n$ cases has required us to generalize finite-difference schemes previously employed successfully in the literature for Burgers-related stochastic equations. We expect these generalizations to be of potential use for the type of higher-order equations that we presently study. Likewise, we hypothesize the potential experimental observation of the scaling behavior that has been elucidated, perhaps in nanofluid (where thermal, or external flux fluctuations are comparatively more relevant) and/or in active matter systems.



\section*{Acknowledgements}
This work has been partially supported by Ministerio de Ciencia e Innovaci\'on (Spain), by Agencia Estatal de Investigaci\'on (AEI, Spain, 10.13039/501100011033), and by European Regional Development Fund (ERDF, A way of making Europe) through Grants No.\ PGC2018-094763-B-I00, No.\ PID2021-123969NB-I00, and No.\ PID2024-159024NB-C21, and by Comunidad de Madrid (Spain) under the Multiannual Agreement EPUC3M23 with UC3M in the line of Excellence of University Professors, in the context of the V Plan Regional de Investigaci\'on Cient\'{\i}fica e Innovaci\'on Tecnol\'ogica (PRICIT). P.\ G.\ and E.\ R.-F.\ acknowledge financial support through contracts No.\ 2022/197 and No.\ 2022/168, respectively, under the EPUC3M23 line. E.\ R.-F.\ also acknowledges financial support from Universidad Carlos III de Madrid through the Margarita Salas program.

\appendix


\section{Renormalization group details}\label{app:1}

The vertex renormalization results obtained for $n=1, 2$, and 3 --- see Section \ref{sec:analytic} --- read

\begin{eqnarray}
    \Gamma_1 &=& \int^> \frac{-\lambda^2 D_1 k^2 \pi }{20q^2(B_1q^2+A_1)^3} \ dq, \\
    \Gamma_2 &=& \int^> -\frac{3 \lambda^2 D_2 \pi (ik)^3}{4B_2^3 q^6} \ dq, \\
    \Gamma_3 &=& \int^> \frac{\lambda^2 D_3 \pi q^2 k^4}{32 (B_3q^2+A_3)^5} \left(34 B_3^2 q^4 + \right. \nonumber \\
     & & 47 A_3 B_3 q^2 + 7 A_3^2 +  18 B_3^3 q^6 + \nonumber \\
     & & \left. 8 B_3^2 A_3 q^4 + 13 B_3 A_3^2 q^2 + 65 A_3^3 \right) \ dq.
\end{eqnarray}


\section{Details on the numerical schemes}\label{app:3}

We begin by describing the method employed for our numerical simulations of Eq.\ \eqref{eq:n} for $n=2$, as it happens to be free from the instabilities affecting the $n=1$ and 3 cases, see  \ref{app:2}. Thus, for $n=2$, Eq.\ \eqref{eq:n} is efficiently simulated through a pseudo-spectral method in which the linear part is solved exactly and the non-linear part is solved numerically by means of an integrating factor, as described in Ref.\ \cite{Gallego07}. We combine the integrating factor with the four-step Adams-Bashforth and three-step Adams-Moulton methods as predictor and corrector methods respectively, as proposed in Ref.\ \cite{Gallego11}.

The space dependence of the remaining $n=1$ and 3 cases is addressed in both cases through finite differences schemes. In both cases, the linear term is implemented through a 9-point centered stencil obtained from Ref.\ \cite{Fornberg88}. This provides $O(\Delta x^8)$ order of accuracy. For the non-linear term, a different discretization is employed for each value of $n$, as decribed below.

For $n=1$, the non-linear term has been discretized by writing the centered first-order derivative of $\phi\partial_x \phi$ and using for the latter the scheme proposed in Ref.\ \cite{Sasamoto09}, which leads to
\begin{align}
    &\partial_x (\phi_i\partial_x \phi_i) \approx \nonumber\\
    &\frac{1}{6\Delta x^2}(\phi_{i+1}\phi_{i+2}-\phi_i\phi_{i+1}-\phi_{i-1}\phi_i+\phi_{i-2}\phi_{i-1} \label{eq:n=1num}\\
    &+\phi_{i+2}^2-2\phi_i^2+\phi_{i-2}^2) . \nonumber
\end{align}

For $n=3$, we design the numerical scheme under the same principle as for $n=1$, by computing the average of three different schemes (backward, centered, and forward) of $\phi\partial_x^3 \phi$ at the same point, and then evaluating its derivative, namely,
\begin{align}
    &\phi_i\partial_x^3 \phi_i\approx\Psi_i=\frac{1}{3 \Delta x}[\phi_{i-1}(-\phi_{i-1}+3\phi_i-3\phi_{i+1}+\phi_{i+2}) \nonumber\\
    &+\phi_i(-\frac{1}{2}\phi_{i-2}+\phi_{i-1}-\phi_{i+1}+\frac{1}{2}\phi_{i+2}) \label{eq:n=3num}\\
    &+\phi_{i+1}(-\phi_{i-2}+3\phi_{i-1}-3\phi_i+\phi_{i+1})] , \nonumber
\end{align}
so that
\begin{equation}
    \partial_x (\phi_i\partial_x^3 \phi_i) \approx \frac{\Psi_{i+1}-\Psi_{i-1}}{2\Delta x} .
\end{equation} 
Both for $n=1$ and 3, the time integration has been implemented through an Euler implicit scheme whose solutions were found iteratively using a Newton-Raphson algorithm.

\section{Stability issues}\label{app:2}

Equation \eqref{eq:n} has been found to be problematic to integrate numerically for odd values of $n$, producing unstable numerical solutions. The reason is the explicit dependence of the non-linear term on the value of $\phi$ when it is accompanied by an even-order derivative, just as in the second term on the righ-hand side of Eq.\ \eqref{sec:deterministic:eq:nlnum} for odd $n$. We discuss each value of $n$ separately.
\\
\\
\noindent $n=1$: 
The $n=1$ case of Eq.\ \eqref{eq:n} is very similar to the noisy Kuramoto-Sivashinsky (nKS) equation (see e.g.\ Ref.\ \cite{Rodriguez-Fernandez2021} and other therein), which is well-known to produce stable solutions at long times in spite of containing a linearly unstable term. Indeed, by expanding the nonlinear term as in Eq.\ \eqref{sec:deterministic:eq:nlnum}, Eq.\ \eqref{eq:n} with $n=1$ reads
\begin{equation}\label{eq:n1b}
   \partial_t \phi=- B_1 \partial_x^4 \phi + \lambda_1 (\partial_x\phi)^2+ \lambda_1 \phi\partial_{x}^{2}\phi + \eta(x,t),
\end{equation}
to be compared with the nKS equation for a scalar field $h(x,t)$, which reads
\begin{equation}\label{eq:nks}
   \partial_t h=- B \partial_x^4 h + \lambda (\partial_x h)^2- \nu \partial_{x}^{2} h + \eta(x,t),
\end{equation}
for $B,\nu>0$. The morphological instability induced by the backward diffusion operator $- \nu \partial_{x}^{2} h$ of the nKS equation is approximated by the $\lambda_1 \phi\partial_{x}^{2}\phi$ term in Eq.\ \eqref{eq:n1b} for $\phi>0$ ($<0$) if $\lambda_1<0$ ($>0$) in regions where the space variation of $\phi(x)$ is relatively slow, e.g.\ near local extrema. This can be further confirmed by linearizing Eq.\ \eqref{eq:n1b} around a constant solution for $\phi(x,t)$.
We decided to avoid the difficulty of dealing with this instability by only considering positive solutions for $\phi$, for $\lambda_1>0$. This can be achieved throughout a sizeable region of parameter space, within which scaling behavior becomes parameter-independent.
\\
\\
\noindent $n=3$: 
A similar discussion can be made for Eq.\ \eqref{eq:n} with $n=3$, which, after expanding its nonlinearity, becomes
\begin{equation}\label{eq:n3b}
   \partial_t \phi=- B_3 \partial_x^4 \phi + \lambda_3 (\partial_x\phi)(\partial_x^3\phi)+ \lambda_3 \phi\partial_{x}^{4}\phi + \eta(x,t).
\end{equation}
In this case, there is a backward (linearly unstable) biharmonic term in the equation for $\phi>0$ ($<0$) if $\lambda_3>0$ ($<0$). This term actually competes with the operator with coefficient $B_3$, so that stability is ensured by considering positive solutions such that $\phi<B_3/\lambda_3$ for $\lambda_3>0$.

%

\end{document}